\documentclass[12pt,a4paper]{article}
\usepackage{amsfonts}
\usepackage{epsfig}
\newcommand{\RR}{\mathbb{R}}

\title{Performance of a new invariants method on homogeneous and non-homogeneous quartet trees\footnote{To appear in Molecular Biology and Evolution}}
\author{M. Casanellas
\and J. Fern\'{a}ndez-S\'{a}nchez
\\
\centerline{Dpt. Matem\`{a}tica Aplicada I} \\
\centerline{Universitat Polit\`{e}cnica de Catalunya, Barcelona, Spain} \\
}

\date{}

\begin{document}

\maketitle

\begin{abstract}
An attempt to use phylogenetic invariants for tree reconstruction was made at the end of the 80s and the beginning of the 90s by several authors (the initial idea due to Lake \cite{Lake1987} and Cavender and Felsenstein
\cite{Cavender87}). However, the efficiency of methods based on invariants is still in doubt (\cite{Huelsenbeck1995},
\cite{JinNei90}), probably because these methods only used few generators of the set of phylogenetic invariants.
The method studied in this
paper was first introduced in \cite{CGS} and it is the first
method based on invariants that uses the \emph{whole} set of
generators for DNA data. The simulation studies performed in this paper prove
that it is a very competitive and highly efficient phylogenetic
reconstruction method, especially for non-homogeneous models on phylogenetic trees.
\end{abstract}

\section*{Introduction}
Since the introduction of phylogenetic invariants by Cavender and
Felsenstein  \cite{Cavender87}, Lake \cite{Lake1987} and Evans and Speed \cite{Evans1993}
several attempts to give a generating set of polynomial phylogenetic invariants have been
made (see for example
\cite{Steel1993},
\cite{FS1995}) but it has not been until recently that algebraic
geometers have managed to find them all \cite{Allman2004b},
\cite{Sturmfels2005}, \cite{CS}.
Methods based on invariants
have already proved to be useful in comparative genomics \cite{Sankoff1999}). However,
a perception seems to have developed that invariants are inefficient, in the technical sense of requiring long sequences to correctly infer a phylogenetic tree, cf. \cite{JinNei90},
\cite{Huelsenbeck1995}. While \cite{Huelsenbeck1995} showed the inefficiency for the use of Lake's invariants alone, no method using all invariants had been proposed at that point, and the question for invariants-based methods in general was never investigated. Note that Lake's method of invariants only used two
phylogenetic invariants of degree one among the 795 generators of
the set of polynomial invariants of a quartet tree for the Kimura
2-parameter model \cite{Garciaweb}. But as Felsenstein
explained, invariants are worth more attention for \emph{what they
might lead to in the future} \cite{Felsenstein2003}.  This
future may be soon, since the studies of this paper show a method based on invariants which is indeed promising. Recently, other methods based on a large set of invariants have also been considered \cite{Eriksson05}, \cite{Rock06}.

Phylogenetic invariants are relationships satisfied by the expected
pattern frequencies occurring in sequences evolving along a given tree topology $T$ under an evolutionary model. More precisely, if \textbf{t} is the set of $d$ model parameters on $T$ and
$p_{\alpha}(\mathbf{t})$ is the probability of observing the pattern
$\alpha$ at the leaves of $T$, by letting  $\mathbf{t}$ vary on
an open subset of $\RR^d$, the probability vector $p(\mathbf{t})=(p_{AA \dots A}(\mathbf{t}),
\dots, p_{TT \dots T}(\mathbf{t}))$ defines a subset $S_T$
of dimension $\leq d$ of $\RR^{4^n}$.  A \emph{phylogenetic invariant} is a real-valued continuous function $f(x)$ on $\RR^{4^n}$ such
that $f(p)=0$ for any $p\in S_T$, but not for all the points on the
subset $S_{T'}$ determined by another tree topology $T'$.
Essentially, the equations $f(p)=0$ are satisfied for pattern frequencies arising from any model parameters on a fixed tree, so they might be used
for recovering the tree topology.

In practice, the vector of observed pattern frequencies $\hat{p}$
obtained from an alignment of sequences for $n$ taxa with enough data, should
approximate $p(\mathbf{t})$ for some set of parameters
$\mathbf{t}$ on a tree topology $T$. In other words, $\hat{p}$
should be a point close to the subset $S_T$ so, if $f$ is an
invariant for the topology $T$, one should have that $f(\hat{p})$
is very close to 0.
As the tree topology is identifiable via invariant-based methods \cite{Allman2006}, using phylogenetic invariants for tree reconstruction is a consistent
method (see \cite{HL}, \cite{Felsenstein2003}). A practical introduction to the theory of invariants can be found in the book of J. Felsenstein \cite[chapter 22]{Felsenstein2003}, whereas the book \cite{ASCB2005} provides a beautiful insight into the applications of algebraic statistics (and in particular polynomial phylogenetic invariants) to computational biology.

There are two major motivations for using phylogenetic invariants
in tree reconstruction. One of them is the prohibitive
computational expense of a full maximum likelihood estimation of a
tree, its edge lengths, a base distribution, and a rate matrix. The other is that the evolutionary
models underlying the theory of invariants allow for non-homogeneous mutation. Indeed, it
is known that, for some biological data sets, different rate matrices
should be allowed in different lineages. Thus it is essential to
have at our disposal phylogenetic methods for reconstructing trees
admitting non-ho\-mo\-ge\-neous models \cite{YY99}, \cite{Galtier1998}.

In this paper, a phylogenetic reconstruction method that uses polynomial phylogenetic invariants  (introduced in \cite{CGS})  is studied and tested for quartet trees
evolving under the Kimura 3-pa\-ra\-me\-ter model of nucleotide substitution \cite{Kimura1981}. Actually, we consider an \emph{algebraic} Kimura model: the parameters of the model are the entries of the substitution matrices on the edges (and not a single rate matrix together with edge lengths). Hence the model is non-homogenous ---because it allows different rate matrices among the edges--- but it is stationary (and the distribution of the bases is uniform), and we always assume
that all sites are independent and identically distributed (i.i.d.
hypotheses).
We performed simulation studies to
test its efficiency. One of our approaches to evaluate the performance and efficiency of the
method is taken from Huelsenbeck \cite{Huelsenbeck1995} so that a
large portion of the tree space is examined to get a general idea
of how the algorithm performs. We present the results obtained for
sequences of length 100 up to 10000.

We also checked the performance of the method on simulated data from non-ho\-mo\-ge\-neous models by carrying out a comparison to Neigh\-bor-Joi\-ning algorithm \cite{Saitou1987}, maximum likelihood algorithm \cite{Felsenstein1981}, and an
algorithm for a non-ho\-mo\-ge\-neous model from \texttt{PAML} \cite{Yang1997} for
sequences generated under a Kimura 2-parameter model  \cite{Kimura1980} and different
rate matrices along different tree branches.

\section*{Results}
\subsection*{Homogeneous data}
\begin{figure*}[p]
\begin{center}
\includegraphics[width=16cm,height=14cm]{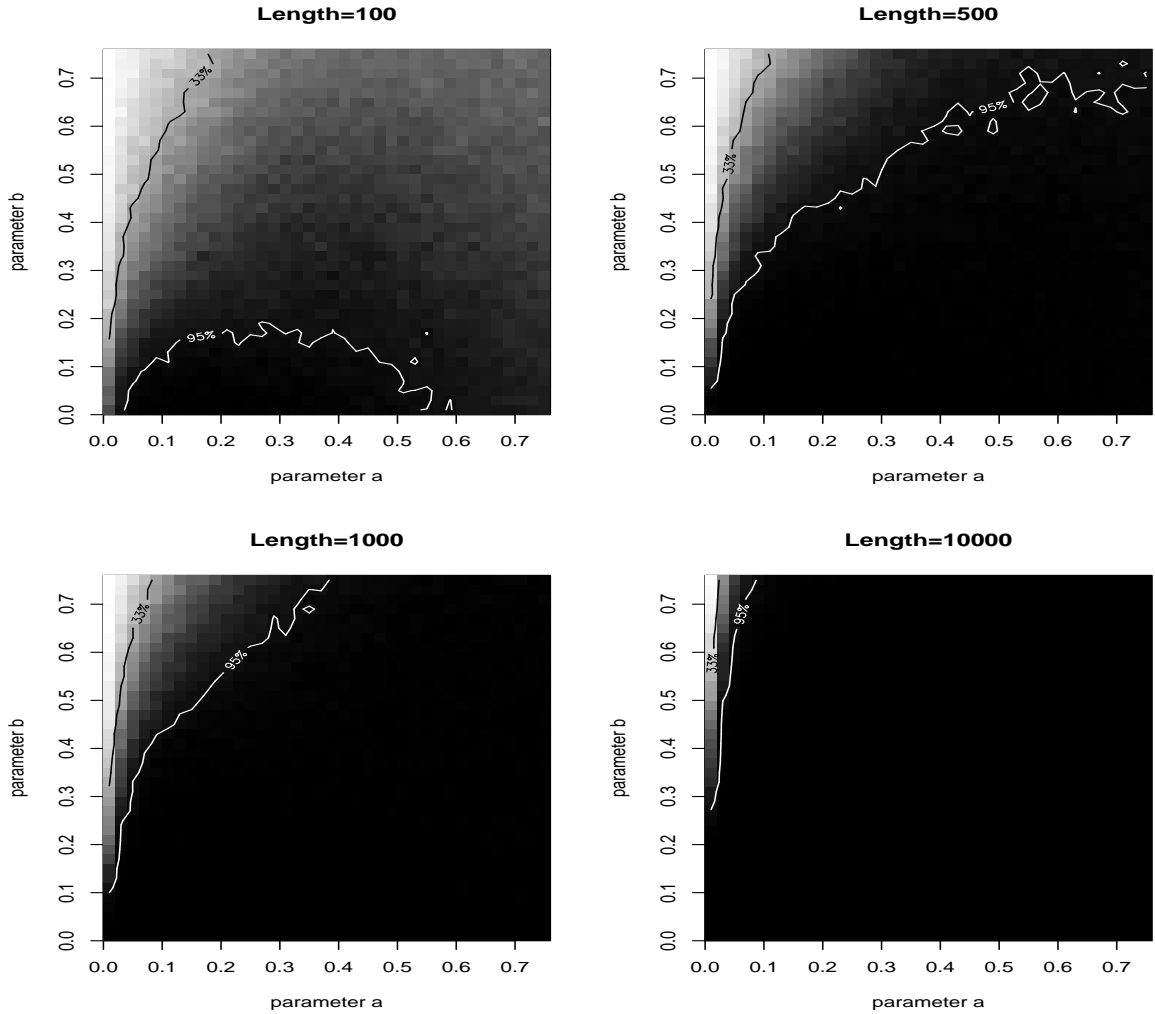}
\end{center}
\caption{\label{fig:inv5} \small The graphics above represent the probability of
reconstructing the correct tree in the parameter space (see Figure I in the \textit{Supplementary Material}). Black areas correspond to couples
$(a,b)$ for which the tree was correctly estimated, while white
regions correspond to couples $(a,b)$ for which the tree is never
estimated; grey tones indicate areas of intermediate probability.
The $95\%$ isocline is drawn in white, while the $33\%$ is drawn
in black. The four graphics above
show the results obtained under the homogeneous Kimura 3-parameter model when using a rate
matrix with parameters $\gamma=0.1, \alpha=3.0, \beta= 0.5$ (hence a 5:1 transition:transversion
bias).}
\end{figure*}

The performance of the invariants method studied here on homogeneous Kimura models can be seen in the figure \ref{fig:inv5}.  Using the approach of J.P. Huelsenbeck in \cite{Huelsenbeck1995} for quartet trees, we considered two branch-length parameters $a,b$ and simulated data for each pair of lengths. Parameter $a$ assigns the branch length to the
internal branch and two opposite peripheral branches, and parameter $b$ assigns the branch length to the
two remaining branches. Parameters $a$ and $b$ were varied from 0.01 to 0.75 in increments of 0.02, and 1000 alignments were simulated for each couple $(a,b)$ (see figures 2 and 3 in \cite{Huelsenbeck1995}, or figure I in the \emph{Supplementary Material} for a clear picture of this space of parameters). The simulated trees evolve under the Kimura 3-parameter model \cite{Kimura1981} with a fixed rate matrix of the form

\begin{tiny}
\begin{eqnarray*}
&  \, \, A \, \,  \, \, C \, \,\, \,  G \, \,\, \, T & \\
\begin{array}{c} A \\ C \\ G \\
T \end{array} & \left( \begin{array}{cccc} \cdot & \gamma & \alpha & \beta \\
\gamma & \cdot & \beta & \alpha \\  \alpha & \beta & \cdot & \gamma \\ \beta & \alpha & \gamma
& \cdot \\ \end{array} \right)
\end{eqnarray*}
\end{tiny}

\noindent
along the tree ($\cdot=-\alpha -\beta -\gamma $). Figure \ref{fig:inv5} shows the efficiency of the method considered in this paper for a rate matrix with parameters $\gamma=0.1, \alpha=3.0, \beta=0.5$ (hence, a 5:1 transition:transversion bias) and for nucleotide sequences of lengths 100, 500, 1000 and 10000. See also figures III and IV of the \textit{Supplementary Material} for studies performed with other rate matrices.

This figure is to be compared with those shown in Figure A2 of  \cite{Huelsenbeck1995} (corresponding to phylogenetic inference for sequences generated
under a Kimura 2-pa\-ra\-me\-ter model of substitution \cite{Kimura1980}
with 5:1 transition:transversion bias). Though this is of course a
biased comparison because our method admits non-homogeneous data and Kimura 3-parameter model, it is worth noticing that even on data from homogeneous simulations our method outperforms many of the methods considered there. In particular, it is clearly better than Lake's invariant method (which is not surprising because Lake's method only used linear invariants). At least for sequences of length 500 or larger, our method performs better than neighbor-joining
---referred as Minimum Evolution (Kimura$_1$) in figure A2 of Huelsenbeck's paper. Notice also that the shape of the
$95\%$-isocline for lengths around 500 or larger is quite
different from the corresponding shapes in the methods tested by
Huelsenbeck (see his figure 7): for large values of $a$ (near 0.7), the performance
of our method of invariants does not drop drastically (as it does for most methods considered there). Therefore for these values our invariants method outperforms all the methods studied in \cite{Huelsenbeck1995}. As it can be seen in figures III and IV in the \textit{Supplementary Material}, it seems that the method performs better for small transition: transversion ratios.

For length 1000, the efficiency of the invariants method considered here is
similar to that obtained for lengths $\geq 10000$ in many of
the methods tested in  \cite{Huelsenbeck1995}. From this it can be
inferred that, in order to reconstruct the correct tree, much less data
is needed in the invariants method presented here than in many other
methods (contrary to what was thought until now  \cite{HL}).

\subsection*{Non-homogeneous data}
We tested the invariants method studied in this paper on data simulated according to a non-homogeneous Kimura model by comparing it with other methods.
\subsubsection*{1. Comparison with neighbor-joining}
\begin{figure}[ht]
\begin{center}
\includegraphics[width=8cm,height=7cm]{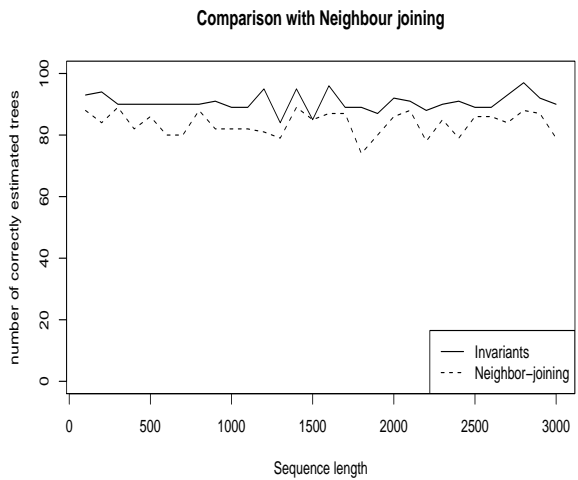}
\end{center}
\caption{\label{fig:usvsnj} \small Performance of neighbor-joining and the invariants method on non-homogeneous data for sequence length varying from 100 to 3000 in intervals of 100. For each length, 100 sets of data were generated. The non-homogeneous tree used for simulations is described in figure \ref{fig:tree5}.}
\end{figure}

\begin{figure}[ht]
\begin{center}
\includegraphics[width=5cm,height=4cm]{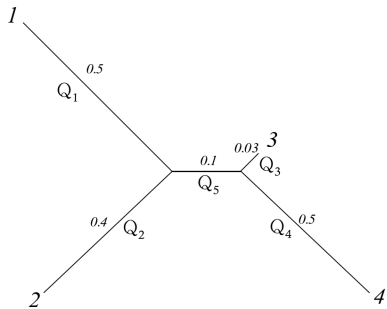}
\end{center}
\caption{\label{fig:tree5} \small A non-homogeneous tree used for simulations. The numbers labelling each edge
correspond to branch lengths (i.e. expected percent change between the two taxa on the edge). (1) In comparing our
method with neighbor-joining, the parameters $\alpha_i,\beta_i, \gamma_i$ in the Kimura 3-parameter rate matrices
$Q_i$ were chosen as: $\gamma_1=1,\alpha_1=4, \beta_1=1$, $\gamma_2=5,\alpha_2=14, \beta_2=3$
,$\gamma_3=4,\alpha_3=15, \beta_3=3$, $\gamma_4=2,\alpha_4=6, \beta_4=2$, $\gamma_5=2,\alpha_5=3, \beta_5=1$. (2)
In comparing our method with maximum likelihood algorithm based on the Kimura 2-parameter model, we chose
$\gamma=\beta=1$ in all rate matrices whereas parameter $\alpha$ was set as follows: in $Q_1$, $\alpha=4$, in
$Q_2$, $Q_4$ and $Q_5$, $\alpha=3+\varepsilon^2$, in $Q_3$, $\alpha=3+\varepsilon$. When $\varepsilon=1$ the tree
is homogeneous and we studied the efficiency of three methods as $\varepsilon$ increases up to 9.}
\end{figure}

First of all we compared the performance of the invariants method presented here with
neighbor-joining (the algorithm of \cite{Saitou1987}) using Kimura  3-parameter distance \cite{Kimura1981} . As it can be seen in figure \ref{fig:usvsnj}, considering certain non-ho\-mo\-ge\-neous sets of simulated data, we found that the invariants method is more efficient than neigh\-bor-joi\-ning.
Indeed, we simulated data on an unrooted quartet tree evolving under the Kimura 3-parameter model where different
rate matrices at each edge were chosen (see figure \ref{fig:tree5}(1)) and we studied the efficiency of the method
when varying the length of nucleotide sequences. In this case, the mean of correctly reconstructed trees for the
invariants method is 90.2$\%$ whereas the mean for neighbor-joining is 84$\%$. It is worth pointing out that the
use of Kimura distance is still justified in the non-homogeneous setting, so that it is fair to compare both
methods. For this particular tree, the maximum likelihood algorithm for the Kimura 3-parameter model reconstructed
the tree correctly almost all times, so we do not include the results here.

\subsubsection*{2. Comparison with maximum likelihood}
We compared the invariants method with two versions of maximum-likelihood: the usual maximum likelihood for Kimura
2-parameter \cite{Kimura1980} and non-homogeneous maximum likelihood method developed in the package \texttt{PAML}
\cite{Yang1997} for Kimura 2-parameter model (namely, the option \emph{nhomo}). This last method allows different
tran\-si\-tion/{\allowbreak}trans\-ver\-sion ratio in different tree  branches. To perform this comparison we
simulated data according to the tree in figure \ref{fig:tree5}(2) evolving under the Kimura 2-parameter model.
When $\varepsilon=1$ the tree is homogeneous and we studied the efficiency of the three methods as $\varepsilon$
increases up to 9.
\begin{figure}[ht]
\centering
\includegraphics[width=7cm,height=6cm]{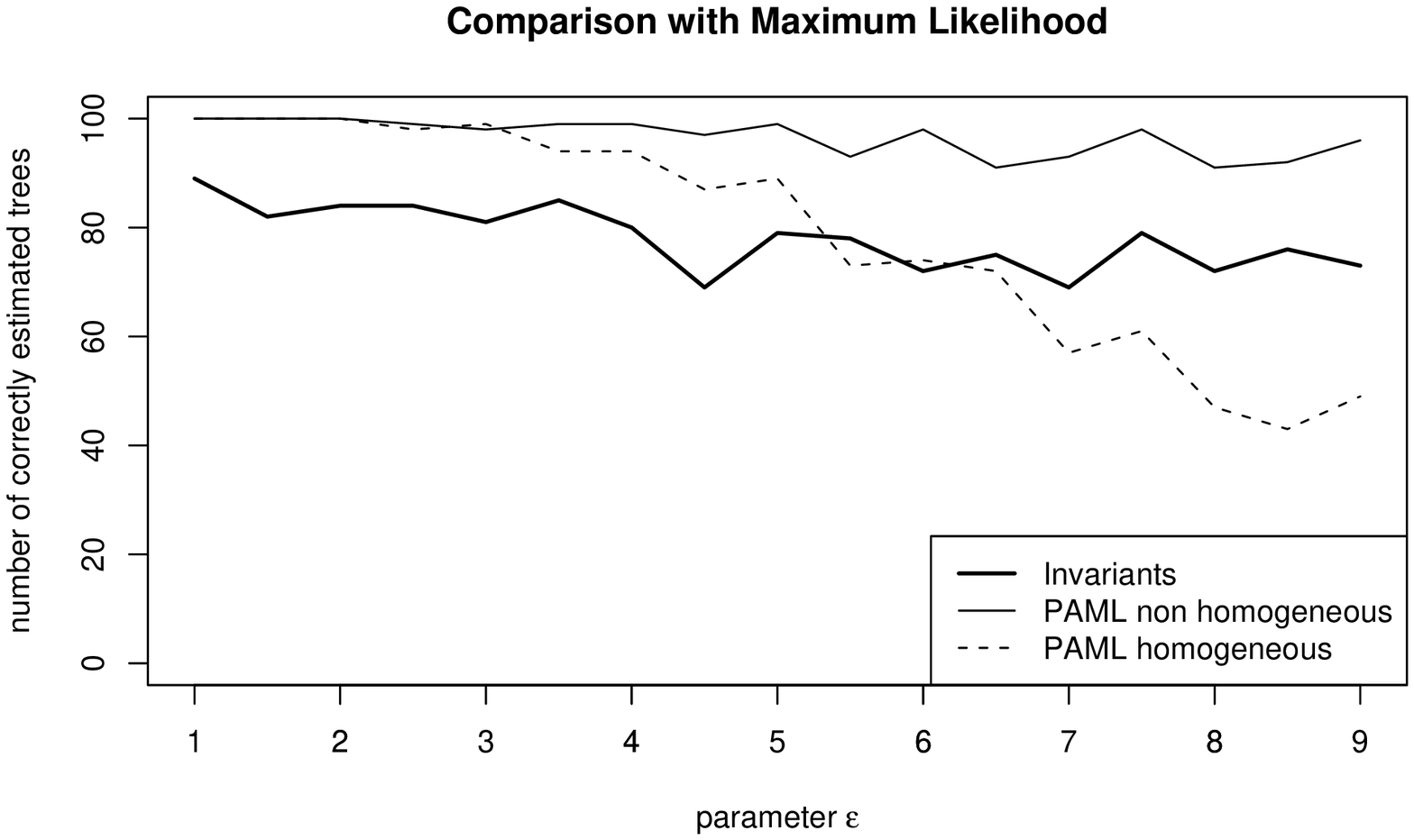}
\caption{\small
The effect of rate heterogeneity among lineages on
method performance. The simulated nucleotide sequences have length 1000 and evolve under the non-homogeneous Kimura 2-parameter model as described in figure \ref{fig:tree5} (homogeneous for $\varepsilon=1$). The trees were reconstructed with our method (\texttt{invariants}),
with \texttt{PAML} under a non-homogeneous method (\texttt{PAML non-homogeneous}) and the usual \texttt{PAML} maximum likelihood homogeneous (\texttt{PAML homogeneous}). For each $\varepsilon$ 100 sets of data were generated.} \label{fig:usvsml}
\end{figure}
Figure \ref{fig:usvsml} summarizes the results relative to the
comparison between our method, the non-homogeneous method in
\texttt{PAML} for Kimura 2-parameter mod\-el (option \texttt{nhomo=2} in the baseml control file)
and the maximum likelihood homogeneous in \texttt{PAML} for Kimura 2-parameter (option \texttt{nhomo=0}). It shows the percentage of
correctly reconstructed trees for each value of parameter
$\varepsilon$.
As expected, the homogeneous maximum-likelihood algorithm becomes less efficient as $\varepsilon$ increases. The use of a method assuming homogeneity on data produced in accord with a non-homogeneous model is, of course, not recommended as such model misspecification is known to lead to unreliable results.  Already for $\varepsilon=5$, the invariants method overtakes the homogeneous maximum likelihood algorithm. As it is deduced from figure \ref{fig:usvsml}, our method performs worse than the non-homogeneous algorithm in \texttt{PAML}. However, it is worth noticing that in this test we generated data according to Kimura 2-parameter model and the invariants method presented here was developed under Kimura 3-parameter model. Similar results are obtained when the data evolve under Kimura 3-parameter model (see figure II in the \textit{Supplementary Material}).

\section*{Methods}
The phylogenetic reconstruction method used in this paper is based
on phylogenetic invariants and was first introduced by the first
author and L.D. Garcia and S. Sullivant in \cite{CGS}.

The taxa are given
by an alignment of $n$ DNA sequences of length $N$. A Markov process along an unrooted binary tree  of $n$ species is assumed and we consider that
all sites are independently and identically distributed (i.i.d.
hypotheses).
The parameters of the model we are considering are the
entries of the substitution matrices of a Kimura 3-parameter model \cite{Kimura1981}. (we should rather speak of an
\emph{algebraic} Kimura 3-parameter model, according to the book
\cite[chapters 1 and 4]{ASCB2005}). Note that in the usual Kimura 3-parameter model, a rate matrix $Q$ is fixed and common
to the whole tree and the substitution matrix is the exponential $e^{Qt_i}$ (for
some parameter $t_i$ representing time). However, in our method we
do not make use of rate matrices $Q$ ---we only use the
substitution matrices--- so, as we will see later, the rate
matrices might vary among the edges.  It is
worth pointing out that, as we consider a Kimura 3-parameter model
along all branches, the uniform distribution of base composition
holds for the whole tree (stationarity hypothesis).

\subsection*{Phylogenetic invariants}
Sturmfels and Sullivant \cite{Sturmfels2005} gave an explicit
description of the generators of the set of polynomial phylogenetic
invariants $I(T)$ for an arbitrary tree evolving under a
\emph{group-based model}. For the Kimura 3-pa\-ra\-me\-ter model on an
unrooted 4-taxa tree the ideal of phylogenetic invariants has 8002
minimal generators (see the webpage \cite{Garciaweb}, \cite{CGS}
and the discussion in \cite{Sturmfels2005}, section 7 to see why a
smaller subset of invariants does not suffice). According to the
results in \cite{Sturmfels2005}, we produced this generating set
for an unrooted tree with 4 leaves under the Kimura 3-pa\-ra\-me\-ter
model.
This requires doing a Fourier transform (or Hadamard
conjugation) on the vector of probabilities $(p_{A \dots A}, \dots,
 p_{T \dots T})$ and the phylogenetic invariants are then described as binomials in the Fourier coordinates.

\subsection*{Algorithm}
Our tree reconstruction algorithm performs the following tasks.
Given 4 aligned DNA sequences $s_1$, $s_2$, $s_3$, $s_4$, it first
computes the observed relative frequencies of each pattern for the
topology $((s_1, s_2), s_3, s_4)$ on an unrooted quartet tree. Then it transforms these
relative frequencies to Fourier coordinates. From this, we compute
the Fourier coordinates in the other two possible topologies for
unrooted trees with 4 species. We then evaluate all phylogenetic invariants
for the Kimura 3-parameter model in the Fourier coordinates of
each tree topology. We call $s_f^T$ the absolute value of this
evaluation for the polynomial $f$ and tree topology $T$. From
these values $\{ s_f^T\}_{_f}$, we produce a score for each tree
topology $T$, namely $s(T)=\sum_f |s_f^T|$. The algorithm then
chooses the topology that corresponds to the minimum score. The
code was written in \texttt{PERL} and is available upon request. For an alignment of 4
sequences of 1000 nucleotides it takes 0.35s on a single 3.0-Ghz
processor.

\subsection*{Software used}
The simulations of this study were obtained using the program
\texttt{Seq-Gen} v1.3.2
\cite{Rambaut1997}. We used the algorithms implemented in the package \texttt{APE} \cite{APE} v1.8-3 of \texttt{R} \cite{R} v2.1.1 to compute Kimura 3-parameter distance \cite{Kimura1981} and perform neighbor-joining algorithm \cite{Saitou1987}.
The package \texttt{PAML} \cite{Yang1997} was used for phylogenetic inference involving maximum likelihood methods.

\section*{Discussion}
The simulation studies performed in this paper present a very
competitive phylogenetic reconstruction method based on
invariants.
If one
compares our results on the tree space and those of Huelsenbeck
\cite{Huelsenbeck1995} one sees that the method presented here is highly efficient. Note that this might be a biased comparison as in both papers homogeneous models are used for sequence generation and, while he uses homogeneous methods for inference, our invariant method allows non-homogeneous models ---re\-mem\-ber that one generally obtains the best results using the most restricted model that fits the data. There are some limitations of
the tree space study performed here, though. For example, this
tree space does not consider trees where the inner edge is
extremely small or extremely large with respect to the peripheral
branches. As
Huelsenbeck points out, the usefulness of considering this
parameter space can be questioned, but he also gives strong arguments
that convinced us to work in this parameter space. Moreover,
considering the same parameter space as Huelsenbeck allows one to
compare our results to the other methods studied by him.

We would like to comment on the algorithm
presented here and, more generally, on all methods based on
invariants. First of all we need to emphasize that the computation of the
phylogenetic invariants of a given model just need to be computed once, so their computation need not contribute to the running-time of an algorithm using them. Secondly, increasing the size of the
sequences does not drop the computational efficiency of the
algorithm. Indeed, the sequences length only accounts for
computing the relative frequencies of the observed patterns (which
is something that most algorithms based on evolutionary models
must do), but it does not participate in any other part of the
algorithm.

A small comment on the election of the 1-norm: we
performed simulation studies not presented here to prove that the
algorithm performs clearly better with the 1-norm than with the
maximum norm, and slightly better
than with the euclidean norm \footnote{The 1-\emph{norm} of a vector $x=(x_1,\ldots,x_n)$ is $\Vert x \Vert_1=\sum_{i=1}^n|x_i|$, the \emph{euclidean norm} is $\Vert x \Vert_2=\sqrt{\sum_{i=1}^n x_i^2}$ and the \emph{maximum norm} is $\Vert x \Vert_{\infty}=\max_{i} |x_i|$.}. Another consideration that might be
important for the computational efficiency of the method is that,
in Fourier coordinates, the polynomials considered here are
\emph{binomials} and hence they are easy to evaluate at a given
point (so there is no need to worry computationally about the
evaluation of the polynomials). Moreover, as it is proved in
\cite{Sturmfels2005}, these binomials have degree 4 at most so,
again, the computational cost is low. Choosing another generating set of invariants or a different weightening of the polynomials will lead to other results, so this is an issue that should certainly be studied in the future.

We implemented and tested the algorithm presented here only on
4-taxon trees. This seems a limitation of the method but as the
reader may have noticed, the method is universal and could be used
to infer the topology of trees with arbitrary number of taxa.
However, the computational demands of deducing large phylogenies
led us not to develop this algorithm for larger trees. Instead, we
suggest that invariants might be a good starting point for quartet
methods of phylogenetic inference. In this direction, it is also
worth thinking about new tree reconstruction algorithms for
arbitrary taxa based on invariants (this is something the authors
will surely work on in the future).

In this paper we focused on the Kimura 3-parameter evolutionary
model. However, a full generating set of invariants is known for
any group-based model  (\cite{Sturmfels2005}), a large set of
invariants is known for the general Markov model
\cite{Allman2003}, \cite{Allman2004b} and for a strand symmetric model \cite{CS}, and some invariants are already known for certain rate-class models \cite{Allman2006}.  Therefore, the method
presented here can be extrapolated to these evolutionary models
and in the future further models can be considered. In particular, invariants might also perform well for models allowing site-to-site variation.

As we pointed out in the introduction, invariants based methods
focus on recovering the tree topology and not estimating the
parameters. Nevertheless, as Allman and Rhodes say
(\cite{Allman2003}, \cite{Allman2004}), it is fair to think that
phylogenetic invariants may also be useful for parameter recovery.

As we have already mentioned, one of the advantages of the method presented here versus other
methods of phylogenetic reconstruction based on evolutionary
models is that the model considered here is non-ho\-mo\-ge\-neous in the
sense that the rate matrix is allowed to vary along the different
branches of the tree. However, as we are assuming a Kimura model, the base distribution is the same
at all nodes of the tree and so the model is stationary (note that invariants methods based on other evolutionary models would not require stationarity of base distribution). For an
unrooted tree with $n$ taxa, our algebraic Kimura model has
$3(2n-3)$ free parameters, and it is a special case of the
\emph{general Markov model}
which involves $12(2n-3)+3$ parameters. If one considered a Kimura
3-parameter model (not the \emph{algebraic} model considered here)
allowing different rate matrices along the tree one would have
$3(2n-3)+2n-4$ (the extra parameters correspond to time
parameters). Other non-homogeneous models have been considered in
the literature, see for example \cite{Galtier1998} and
\cite{Yang1995}. The emphasis in these two papers is put on the
non-stationarity hypothesis, and the maximum likelihood approach
taken in most non-homogeneous methods makes them computationally
unfeasible.

\subsection*{Supplementary material}
Supplementary material includes figures I, II, III and IV.

\vspace*{2mm}

\textbf{Acknowledgments:} The authors would like to thank B. Sturmfels, L. Pachter and N. Eriksson for useful comments on a draft version of the paper.  We would also like to thank the reviewers for useful comments that improved the paper. Both authors were partially supported by Ministerio de Educaci\'on y Ciencia (Ram\'on y Cajal and Juan de la Cierva programs, respectively), and BFM2003-06001. The second author was also supported by MTM2005-01518.

\bibliographystyle{alpha}
\bibliography{CasanellasFernandez}

\onecolumn
\begin{figure}[p]
\centering
\includegraphics[scale=0.5]{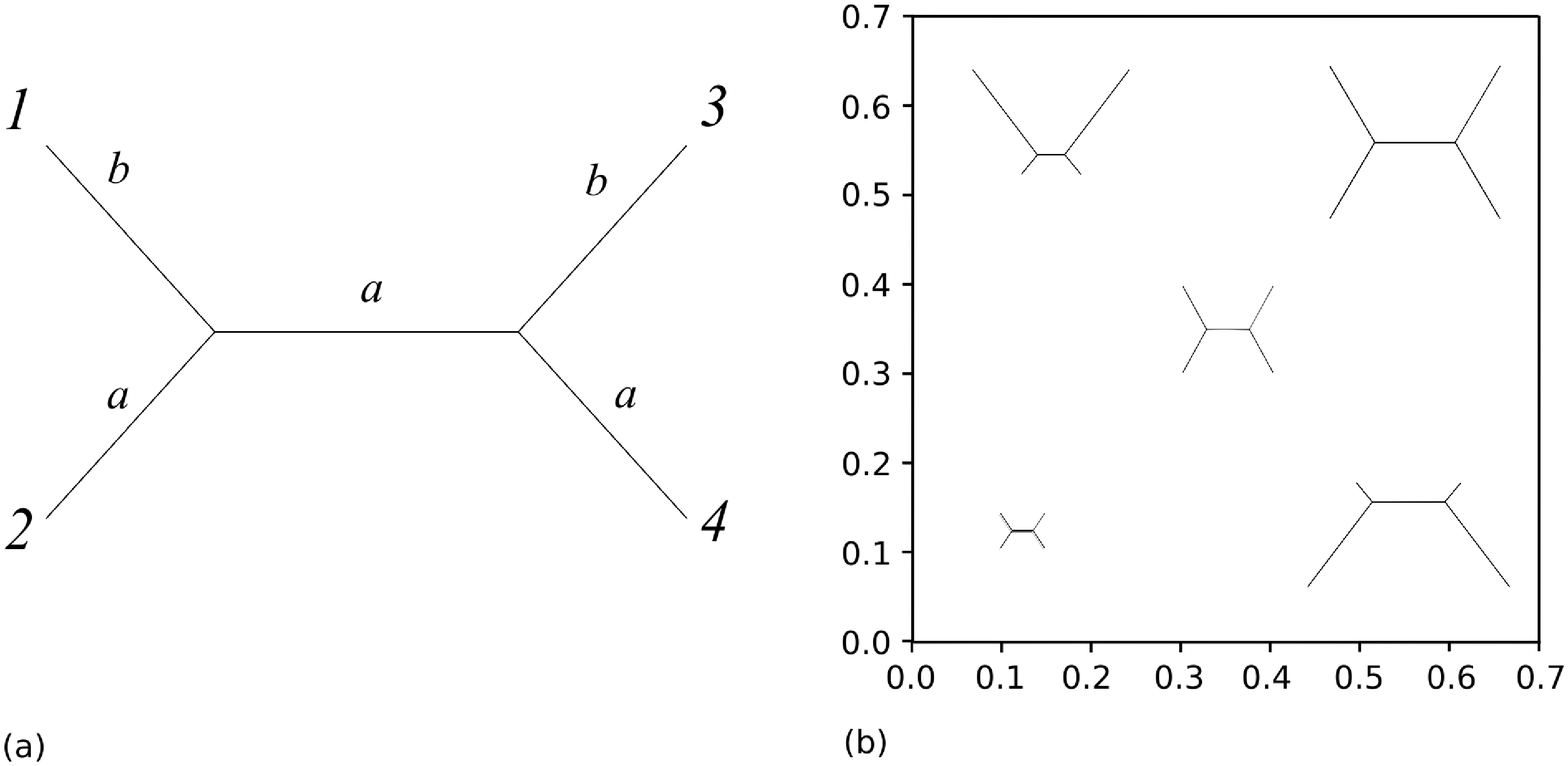}
\caption{\small (a) Unrooted 4-taxa tree. The branches labelled with
the same letter (either $a$ or $b$) have the same branch length. The taxa are named with arabic numbers, $1$ to $4$. (b) The parameter space for the unrooted four-taxon
tree. Parameter $a$ is plotted in the abscissa, while parameter $b$
in the ordinate. The lengths of the branches $a$ and $b$
are varied from $0.01$ to $0.75$ in increments of $2\%$.}
\label{fig:paramsp}
\end{figure}
\begin{figure}[p]
\begin{center}
\includegraphics[scale=0.7]{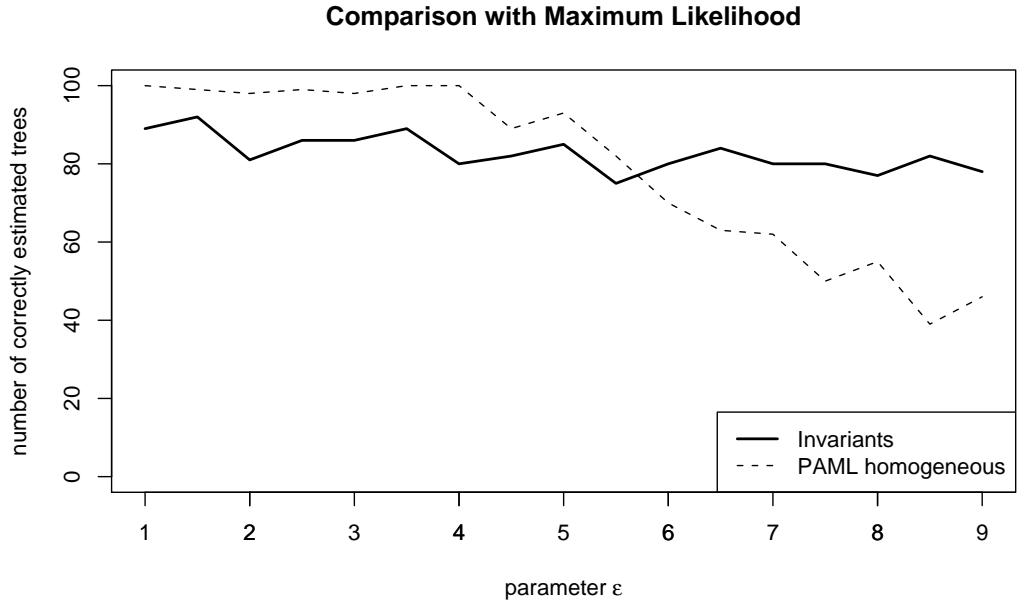}
\end{center}
\caption{\small 
The effect of rate heterogeneity among lineages on
method performance. The simulated nucleotide sequences have length 1000 and evolve under the non-homogeneous Kimura 3-parameter model for the parameters $\gamma=1, \beta=2$ in all rate matrices while parameter $\alpha$ was set as follows: in $Q_1$, $\alpha=4$, in $Q_2$, $Q_4$ and $Q_5$, $\alpha=3+\varepsilon^2$, in $Q_3$, $\alpha=3+\varepsilon$ (see figure 3). The trees were reconstructed with our method (\texttt{invariants}),
and with the usual \texttt{PAML} maximum likelihood homogeneous (\texttt{PAML homogeneous}). For each value of $\varepsilon$, 100 sets of data were
generated.} \label{fig:Kim3}
\end{figure}
\newpage
\begin{figure}[p]
\begin{center}
\includegraphics[width=16cm,height=14cm]{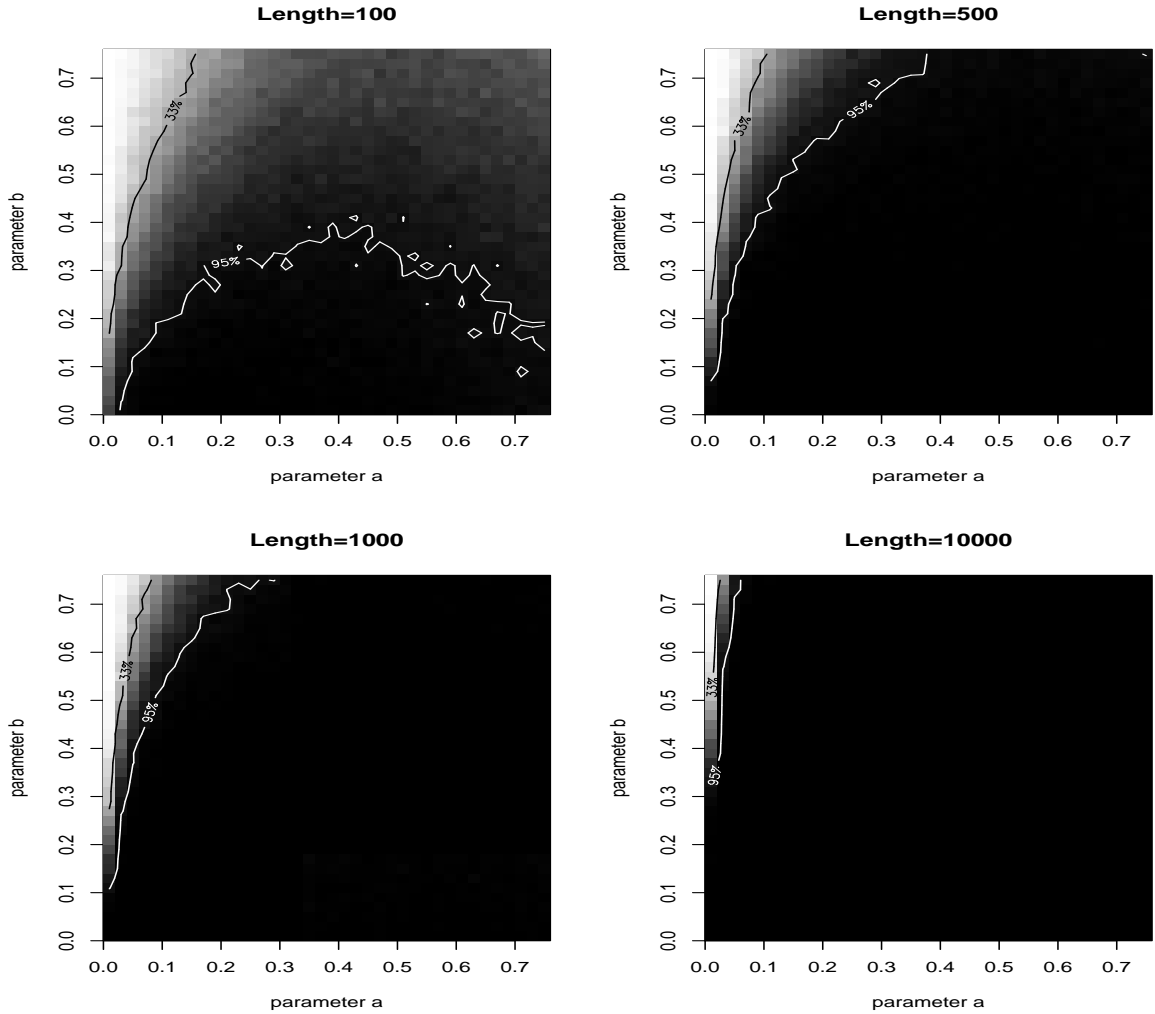}
\end{center}
\caption{\label{fig:inventedparam} \small The graphics above represent the probability of
reconstructing the correct tree in the parameter space (see Figure I). Black areas correspond to couples
$(a,b)$ for which the tree was correctly estimated, while white
regions correspond to couples $(a,b)$ for which the tree is never
estimated; grey tones indicate areas of intermediate probability.
The $95\%$ isocline is drawn in white, while the $33\%$ is drawn
in black. The four graphics above
show the results obtained under the homogeneous Kimura 3-parameter model when using a rate
matrix with parameters $\gamma=0.2, \alpha=0.5, \beta= 0.3$ (hence a 1:1 transition:transversion
bias).}
\end{figure}
\newpage
\begin{figure}[h]
\begin{center}
\includegraphics[width=16cm,height=14cm]{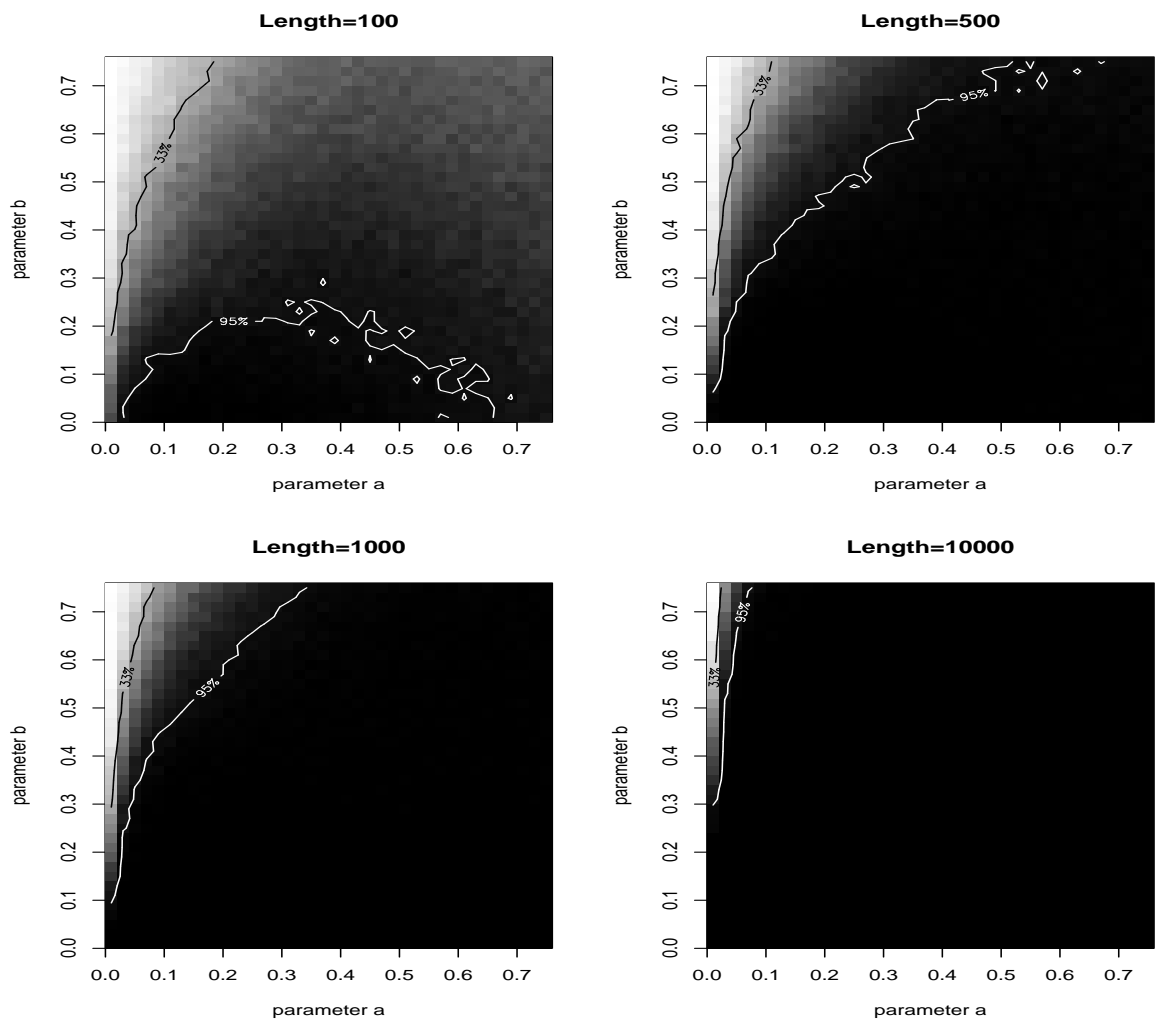}
\end{center}
\caption{\label{fig:realparam}\small The four graphics above show the results obtained in the
four-taxon space when the sequences are generated under the homogeneous Kimura 3-parameter
model of substitution using a rate matrix with parameters $\gamma=0.12, \alpha=0.73, \beta=0.13$ (hence a 2.93:1 transition:transversion bias). This rate matrix corresponds to the maximum likelihood
estimate from an alignment of homologous sequences of eight vertebrates
under a Kimura 3-parameter model.
} 
\end{figure}
\end{document}